# Single crystal growth and superconducting properties of LiFeAs

Bumsung Lee[1], Seunghyun Khim[1], Jung Soo Kim[2], G. R. Stewart[2] and Kee Hoon Kim[1,*]

[1] *CeNSCMR & Department of Physics and Astronomy, Seoul National University, Seoul 151-747, Republic of Korea*

[2] *Department of Physics, University of Florida, Gainesville, Florida 32611-8440, U. S. A.*

**Abstract**

We report the successful growth of high quality single crystals of LiFeAs with lateral sizes up to 5 × 5 mm$^2$ by the Sn-flux method. Electrical resistivity studies reveal that the superconducting onset temperature is 18.2 K with a transition width less than 1.1 K and the ratio of room temperature to residual resistivity is about 24. Bulk superconductivity is supported by perfect shielding in the magnetic susceptibility and a clear jump in the specific heat $C_p$, resulting in $\Delta C_p/T \approx 20.0$ mJ/mol·K$^2$. Upper critical field slopes of $dH_{c2}^{c}/dT \approx -1.39$ and $dH_{c2}^{ab}/dT \approx -2.99$ T/K near $T_c$ predict zero temperature upper critical fields of $H_{c2}^{c}(0) \approx 17.2$ and $H_{c2}^{ab}(0) \approx 36.9$ T and coherence lengths of $\xi_{ab} = 4.4$ and $\xi_c = 2.0$ nm in a single band model. This result points to a modest superconducting anisotropy about 2.3 in LiFeAs.

[*] khkim@phya.snu.ac.kr

**Introduction**

The discovery of superconductivity in LaFeAs(O,F) with a superconducting transition temperature $T_c$ = 26 K [1] has sparked intense research activities on the iron-based superconductors. A large number of quaternary iron-oxypnictides $R$FeAsO ($R$ = Ce, Pr, Nd and Sm), often called '1111' systems, were found to be superconducting with $T_c$ up to 55 K for SmFeAs(O,F) series [2]. Later on, the ternary '122' system, $Ae$Fe$_2$As$_2$ ($Ae$: alkaline-earth), the '11' system, iron-chalcogenides Fe(Se,Te) and another ternary '111' system, $A$FeAs ($A$ = Li, Na) joined to become new iron-based superconductors [3-8].

Among those, the '111' system is one of the most recently discovered materials with relatively poor understanding of its physical properties. A representative compound LiFeAs, of particular focus in this study, showed $T_c \approx$ 18 K [5-7]. It has a tetragonal Cu$_2$Sb-type structure and possesses a single Fe-As tetrahedral layer sandwiched by the double Li layers. This structural characteristic provides a unique opportunity to realize homogeneous Li terminating surface upon cleaving, similar to the case of the Bi-O termination in Bi$_2$Sr$_2$CaCu$_2$O$_{8+\delta}$ [9]. Because various spectroscopic tools including angle-resolved photoemission, scanning tunneling microscopy and optical spectroscopy are quite sensitive to the surface state, the possibility of achieving homogeneous Li termination makes LiFeAs attractive for investigating the intrinsic properties of the iron-pnictide superconductors. In this sense, growing a LiFeAs single crystal with large lateral area is a necessary step toward studying its intrinsic properties.

The flux method has been known to be an effective tool to grow single crystals of the iron-based superconductors such as the '122' and the '11' systems. In particular, the use of Sn-, In- and FeAs-fluxes have been successful in obtaining various '122' single crystals [10-15]. Among these, the Sn-flux showed an advantage over the other methods as high solubility of most metallic elements into Sn makes the growth temperature relatively low [10, 11]. The low growth temperature is an important parameter to obtain large single crystals as the molten flux can be cooled down over a wide temperature window. With this advantage, the Sn-flux method has been quite successful in growing a broad range of large-area, high-quality '122' single crystals, such as electron-doped Sr(Fe,Co)$_2$As$_2$ and un-doped SrFe$_2$As$_2$, EuFe$_2$As$_2$ and CaFe$_2$As$_2$ [11-13] while the inclusion of Sn in Sn-flux grown

BaFe$_2$As$_2$ alters the intrinsic properties such as $T_{SDW}$ in this compound significantly [10]. For the CaFe$_2$As$_2$, quantum oscillation has been even observed, supporting the feasibility of using the Sn-flux to obtain a single crystal with low impurity scattering [13]. All these previous works indicate that the Sn-flux method can be used selectively to grow high quality iron-pnictide crystals, with the caveat that its feasibility should be checked for each case.

For LiFeAs, in particular, the high volatility of Li ions has been an intriguing hurdle in the attempts to grow a stoichiometric single crystal. Even though FeAs self-flux has been reported to work for growing LiFeAs single crystals, a high melting temperature of the FeAs flux ~ 1030 °C, demands an additional technique such as a welding of a metal crucible under high Ar pressure to reduce Li evaporation [16]. A recent report using the Bridgeman technique also employed the welding of a metal crucible to keep Li from evaporation during reaction [17].

In this letter, we show that the Sn-flux method by use of conventional quartz ampoule sealing is an alternative to the above methods to obtain a large-area, high-quality LiFeAs single crystal at much lowered growth temperatures ~ 800 °C. Resistivity, DC magnetic susceptibility and specific heat studies are used to ensure its high quality and bulk superconductivity. The study of upper critical field up to 9 T is used to calculate zero temperature coherence lengths and a moderate superconducting anisotropy of 2.3.

**Experimental**

We employed the Sn-flux method to grow LiFeAs single crystals. Stoichiometric amount of Li, Fe and As were weighed and kept in a pair of alumina crucibles, and Sn was added as a flux with a molar ratio [LiFeAs]:Sn = 1:10. The alumina crucible was put into a quartz ampoule, which was sealed under partial Ar atmosphere (~ 0.7 bar) to minimize Li evaporation during the reaction. To avoid oxidation, all the processes for handling chemicals were performed inside a glove box, where the levels of oxygen and moisture were maintained less than 1 ppm. The sealed quartz ampoule was heated up with a rate of 50 °C/h to be kept at 250 °C for 24 hours and subsequently at 500 °C for 4 hours to fully dissolve Li and As into the Sn flux. Later, the ampoule was heated up to 850 °C to stay for 4 hours and then cooled down to 500 °C with a rate of 3.5 °C/h. At 500 °C, centrifuging was performed to remove the Sn-flux from the crystal surfaces. The crystals thus harvested showed a plate-like shape and the maximum lateral size reached up to 5 × 5 mm$^2$.

The structure of the grown crystals was first characterized by powder X-ray diffraction (XRD) measurements through the $\theta$-$2\theta$ scan method under air. Before performing the measurement, several crystal pieces were ground and mixed with Apiezon N-grease$^{TM}$ to reduce oxidation. Although the intensity was smaller than usual, the peak positions were consistent with those known for a stoichiometric polycrystalline specimen. Moreover, $c$-axis lattice constant was estimated through the $\theta$-$2\theta$ scan in a piece of single crystal with flat $ab$-surface.

The resistivity measurement along the $ab$-plane was performed by the conventional four-probe technique using a resistance bridge (Lakeshore 370) in combination with physical property measurement system (PPMS$^{TM}$, Quantum Design). The silver conductive epoxy used for electrical contact was applied inside a glove box. Moreover, to reduce oxidation effect during the transport experiments or sample mounting, we covered the transport specimen with a low temperature epoxy (Stycast$^{TM}$ 1266) and cured at room temperature. A vibrating sample magnetometer attached to the PPMS was used for DC magnetic susceptibility measurements. Before this measurement, the sample was well covered with Kapton$^{TM}$ tape inside the glove box so that it is not likely to be exposed to air during the mounting process. Zero-field-cooled (ZFC) and field-cooled (FC) measurements were performed at magnetic field 10 Oe applied along the $ab$-plane. The specific heat $C_p$ was measured using the thermal relaxation method. Details for the platform and the measurement technique have been described elsewhere [18]. The platform temperature was measured using a Cernox$^{TM}$ thermometer. Accuracy of the apparatus was checked by measuring $C_p$ of a piece of high purity gold (NIST standard), of which value is close to that of our LiFeAs crystal near the critical temperature. Agreement of the measured data with the literature value [19] was within ±3 % over the whole temperature range investigated. Then, several single crystals with total mass 18.5 mg were mounted on a sapphire disk using GE7031 varnish (whose specific heat addenda contribution is known) in a glove box to seal the samples away from contact with the atmosphere. In order to check whether there was a reaction between the LiFeAs sample and the varnish, the zero field diamagnetic shielding of the specific heat sample was again checked in 10 Oe applied in the *ab*-plane with the result that the transition was broadened. Thus, the specific heat data are on a sample where it is to be expected that the discontinuity at $T_c$ will be reduced vs. that for an ideal sample.

**Results and discussion**

X-ray diffraction results are shown in fig. 1. All the peaks can be indexed with a tetragonal *P*4/*nmm* group, being consistent with previous reports on polycrystalline LiFeAs [5-7] and other impurity phases were not detected. The *c*-axis lattice constant was 6.35 Å, again consistent with the previously reported one in a polycrystalline sample [5-7]. As the inset of fig. 1 illustrates, a typical lateral size of the grown crystal reached as high as $5 \times 5$ mm$^2$.

From a linear extrapolation of normal and superconducting regions in the resistivity curve, as dashed lines in fig. 2 (a), the superconducting onset transition was estimated as $T_{onset}$ = 18.2 K. And, the zero resistivity was found to be realized at 17.1 K, resulting in the transition width of 1.1 K. This transition width is smaller than the values reported so far, which ranged from 2 to 4 K [5, 6, 17, 20]. This observation supports the high quality of LiFeAs studied here. The inset of fig. 2 (a) shows the resistivity in a wide temperature window from 3 to 300 K. As temperature is lowered, the resistivity monotonically decreases without exhibiting any anomaly, indicating absence of a SDW or a structural transition, being consistent with the previous reports [5-7, 17]. The residual resistivity ratio (RRR), defined as the ratio of the resistivity at 300 K and extrapolated residual resistivity at 0 K, is found to be 24 in this crystal. Although this value is smaller than the reported RRR of ~ 46 in a LiFeAs single crystal grown by the Bridgeman technique [17], we note that the RRR = 24 is one of largest ones among the iron-based superconductors; most undoped '122' systems have much smaller RRR, for example, ~ 3 for SrFe$_2$As$_2$ [11] and ~ 6 for BaFe$_2$As$_2$ [21]. To dates, only a limited number of systems such as KFe$_2$As$_2$ [22, 23] series seem to have larger RRR above 24, reaching ~ 1280.

Temperature dependence of DC magnetic susceptibility measured at magnetic field of 10 Oe is shown in fig. 2 (b) for the ZFC and FC cases. The shielding fraction is measured to be around 100 % at 10 K from the ZFC measurements. No magnetic anomaly was observed up to room temperature, indicating the absence of any magnetic ordering in LiFeAs, in contrast to the cases in the other parent compounds of the iron-based superconductors. The inset of fig. 2 (b) shows that the diamagnetic signal starts to appear around 16.8 K, slightly lower than 17.1 K where zero resistivity is realized.

Consistent with the substantial diamagnetic signal in the sample mounted in GE7031 varnish, the $C_p/T$ data in the inset of fig. 2 (c) show a clear jump-like feature. When the normal state $C_p/T$ was fit with the formula $C_p/T = \gamma + \beta T^2$ (solid line) with $\gamma \approx 35.0$ mJ/mol·K$^2$ and $\beta \approx 0.215$ mJ/mol·K$^4$, resulting $\Theta_D \approx 300$ K, the remaining $\Delta C_p/T$ in fig. 2 (c) shows a rather broadened transition feature. From this, the thermodynamic $T_c \approx 16.8$ K can be estimated as the midpoint of the jump-like feature. The linearly extrapolated, dashed lines yielded $\Delta C_p/T \approx 20.0$ mJ/mol·K$^2$, which is comparable to the recently reported value in a self-flux grown crystal while the thermodynamic $T_c \approx 16.8$ K is higher than the reported $T_c \approx 15.4$ K [24]. These $C_p/T$ results clearly support the realization of bulk superconductivity in our Sn-flux grown LiFeAs crystal. Furthermore, we note that the thermodynamic $T_c \approx 16.8$ K estimated from the $C_p/T$ result is rather consistent with the value from the magnetic susceptibility but as is typical is a little lower than the zero resistivity temperature.

We also find that there is a slight feature around 10 K in the specific heat data, which might correspond to a second band gap opening, as similarly observed by Wei *et al*. in a small self-flux grown crystal of ~ 0.5 mg [24]. However, there exist a broadened transition feature in the $\Delta C_p/T$ curve and a large residual $C_p$ at lower temperature, which presumably comes from the non-superconducting fraction of the sample. We believe that this is a degradation effect from the high reactivity of LiFeAs upon its contact with the GE7031 varnish which reduced the superconducting volume fraction in the single crystals used for the $C_p$ measurements, thus increasing the residual $C_p$ and broadening the transition at the thermodynamic $T_c$. Existence of such large residual $\gamma$ as well as an incipient upturn presumably due to a Schottky anomaly at low temperatures [14], prevented any reliable fit to the data using a two band gap model from being made.

The upper critical field $H_{c2}$ was also estimated thorough the temperature dependent resistivity measurements under constant magnetic field applied along the *c*-axis and the *ab*-plane. $T_c$ at each magnetic field is determined by the criterion that 50 % of the normal state resistivity is realized at $T_c$. The determined resistivity traces for $H//c$-axis and $H//ab$-plane are presented in figs. 3 (a) and (b), respectively. While $T_c$ decreases with increasing $H$ as expected, it is observed that the superconductivity is more robust for $H//ab$-plane, as summarized in fig. 4. Both of the $H_{c2}$ curves for each $H$ directions, i.e., $H_{c2}^c$ and $H_{c2}^{ab}$ are rather linear near $T_c$ but show different slopes. The inset of fig. 4 shows the estimated anisotropy of the $H_{c2}$ values, defined as $\gamma \equiv H_{c2}^{ab}/H_{c2}^c$. In this estimation, the $H_{c2}$ values between the measured data points were interpolated to calculate $\gamma$ values at each temperature. At temperatures ~ 15.5 K just below $T_c$, $\gamma$ was 2.3. This value of $\gamma$ near $T_c$ is clearly smaller than the $\gamma$ values of $KFe_2As_2$ and the '1111' system ($\gamma \geq 5$) [22, 25, 26], but comparable to the $\gamma$ of most of the superconducting '122' systems, which lies within ~ 2-3 [10, 27-29].

The $H_{c2}$ anisotropy $\gamma \equiv H_{c2}^{ab}/H_{c2}^c$ is close to being the same as the anisotropy of the penetration depth $\lambda$ and coherence length $\xi$ via the relation $\gamma = \xi^c/\xi^{ab} = \lambda^c/\lambda^{ab} \equiv \gamma_\lambda$. This comparison is particularly valid near $T_c$ where the anisotropic Ginzburg-Landau equations apply and the orbital limiting effects are dominant for determining $H_{c2}$ [30]. In case the orbital limiting is dominant in one main active band, $H_{c2}(0)$ can be further determined by the slope of $H_{c2}$ curve near $T_c$ as $H_{c2}^{orb}(0) = -0.69$ $dH_{c2}/dT|_{T=T_c}T_c$, as predicted by the Werthamer-Helfand-Hohenberg (WHH) formula [31]. In the present LiFeAs single crystal, from the $dH_{c2}/dT|_{T=T_c} \approx -1.39$ and $-2.99$ T/K for $H//c$-axis and $H//ab$-plane, respectively, the orbital limiting fields are estimated to be $H_{c2}^c(0) \approx 17.2$ T and $H_{c2}^{ab}(0) \approx 36.9$ T, resulting in the coherence lengths $\xi_{ab}(0)$ and $\xi_c(0)$ as $\xi_{ab}(0) = 4.4$ and $\xi_c(0) = 2.0$ nm based on the Ginzburg-Landau relations, $H_{c2}^c = \Phi_0/2\pi\xi_{ab}^2(0)$ and $H_{c2}^{ab} = \Phi_0/2\pi\xi_{ab}(0)\xi_c(0)$. Through this relationship, the $H_{c2}$ anisotropy $\gamma \approx 2.3$ near $T_c$ implies that the zero temperature anisotropy for the coherence length and penetration depth is also close to 2.3. Recently, based on the first-principles calculations with GGA approximation and the assumption of isotropic relaxation time and its independence of the velocity of the conduction electrons, Nakamura et al. [32] predicted the $\gamma_\lambda(0)$ for various iron-based superconductors in the superconducting states. According to their results, $\gamma_\lambda(0)$ of $BaFe_2As_2$ and LiFeAs are relatively small, i.e. ~3, while the '1111' systems such as LaFeAsO are much bigger, ~ 10. Therefore, our observation of a moderate $H_{c2}$ anisotropy of $\gamma \sim 2.3$ in LiFeAs is a bit smaller than the band calculation results based on the GGA approximation. In a recent effort to include the effects of electron correlation by incorporating the dynamic mean field theory plus density function calculation

[33], $\gamma_\lambda(0)$ was predicted to become smaller, about ~ 1.5. Our experimental $\gamma$ ~ 2.3 is obviously getting closer to but is still larger than this prediction. It is interesting to note that $\gamma$ ~ 1.2 was observed in the sample grown by the Bridgeman technique [17], indicating $\gamma$ can be also sensitive to the crystal growth method.

Although the above discussions on the $H_{c2}$ anisotropy at zero temperature region are based upon the assumption of the orbital-limiting scenario, it is important to check the other pair breaking mechanism at low temperatures to correctly understand the temperature dependence of $H_{c2}$. In general, magnetic fields can break the Cooper pairs mainly by two processes, i.e., orbital limiting and Pauli limiting. If the orbital limiting is the dominant process in a single active band, $H_{c2}(0)$ as well as the temperature dependence of $H_{c2}$ can be obtained by the WHH formula [31]. On the other hand, if the Pauli limiting is the dominant process, the $H_{c2}(0)$ for a weakly coupled BCS superconductor is simply given as the Pauli limiting field, $H_P = 1.86T_c$ [34], which predicts $H_P \approx 32.9$ T with $T_c$ = 17.7 K in the present LiFeAs single crystal. The location of $H_p$ is indicated in fig. 4 as the dashed line. Further, the predictions for the $H_{c2}(T)$ curves by the WHH formula, upon fitting the $H_{c2}(T)$ near $T_c$, are given as dotted lines in fig. 4. From these results, we can postulate possible scenarios for the behavior of the $H_{c2}$ curves at lower temperature. Since $H_{c2}^{ab}(0) \approx 36.9$ T as predicted by the orbital limiting is only a little higher than $H_P$ = 32.9 T, it is expected that temperature dependence of $H_{c2}^{ab}$ might be governed by the orbital limiting effects except very low temperatures. Similarly, as $H_{c2}^{c}(0) \approx 17.2$ T, $H_{c2}^{c}(T)$ is mainly subject to the orbital limiting down to zero temperature limit. Therefore, the overall temperature dependence of $H_{c2}$ in LiFeAs is likely to be determined by the orbital limiting effects for both $H//c$-axis and $H//ab$-plane. Moreover, rather a good agreement of the $H_{c2}^{ab}(T)$ curve with the WHH prediction near $T_c$ supports that the orbital limiting within one band model is effective in describing $H_{c2}^{ab}(T)$ behavior at high temperatures near $T_c$.

On the other hand, for $H//c$-axis, the $H_{c2}$ curve is linearly increasing with decrease of temperature, clearly showing deviation from the one band, WHH model. The deviation of measured $H_{c2}^{c}$ from the simulated curve starts from ~ 6 T. Since the $H_{c2}^{orb}(0) \approx 17.2$ T for $H//c$-axis is much smaller than $H_P$ = 32.9 T, the $H_{c2}^{c}$ curve is expected to keep increasing linearly without saturation. The linearly increasing $H_{c2}^c(T)$ tendency implies that the transport is determined by at least two bands, particularly for $H//c$-axis. The linearly increasing $H_{c2}^c(T)$ is not unique to LiFeAs but has been found in other iron-based superconductors such as electron doped '122' systems [27-29], which have shown involvement of multiple active bands in their physical properties. Based on these linear increasing of $H_{c2}^c$ and reasonable description with the one band, WHH model for $H_{c2}^{ab}$, it is predicted that the actual anisotropy $\gamma$ at low temperatures will approach 1.5, much lower than the $\gamma \sim 2.3$ determined near $T_c$. The predicted evolution of $\gamma$ with the temperature should then share some similarity with those found in several '122' systems [27-29]. This remains to be confirmed by the upper critical field study at a higher field region.

We note, however, the predicted $H_{c2}^{ab}(0) \approx 36.9$ T and $H_{c2}^c(0) \approx 17.2$ T are relatively low, compared with those of the other compounds that have similar $T_c$. In the single crystal and the thin film forms of Co-doped $SrFe_2As_2$ with $T_c \approx 20$ K, $H_{c2}(0) \approx 50$ T and becomes isotropic at the zero temperature [27, 29]. In the '11' system, an optimally doped Fe(Se,Te) with $T_c \approx 14$ K, lower than that of the LiFeAs system, $H_{c2}(0) \approx 50$ T and again become isotropic at zero temperature limit [35]. The smaller $H_{c2}(0)$ for LiFeAs stems from the smaller slopes of $H_{c2}$ increase near $T_c$, which results in larger coherence lengths than the '11' and '122' systems illustrated above. Therefore, the LiFeAs single crystals grown by the Sn-flux can be a good model system to study intrinsic properties of iron-based superconductors.

**Summary**

A large single crystal of LiFeAs up to 5 x 5 mm$^2$ was successfully grown by the Sn-flux method. Electrical resistivity studies show that the superconducting onset temperature is 18.2 K with a transition width less than 1.1 K and a ratio of room temperature to residual resistivity of about 24, indicating the high quality of the crystal. DC magnetic susceptibility and specific heat measurements clearly show evidence for bulk superconductivity. The upper critical fields measured for $H$ applied along the $c$-axis and the $ab$-plane indicate that a modest superconducting anisotropy of 2.3 realized near $T_c$ can decrease further with decreasing temperatures as a result of interplay of dominant orbital-limiting and two-band features in LiFeAs.

**Acknowledgements**

We thank valuable discussions with Profs. Hidenori Takagi, Jun Sung Kim, Ji Hoon Shim, Ki-Young Choi and Dr. Junichi Yamaura. This study was supported by NRF, Korea through Creative Research Initiatives, NRL (M10600000238) and Basic Science Research (2009-0083512) programs and by MOKE through the Fundamental R&D Program for Core Technology of Materials. BSL is supported by Seoul R&BD (10543). Work at Florida supported by the US Department of Energy, contract no. DE-FG02-86ER45268.

**Figure captions**

Fig. 1 (Color online) X-ray diffraction pattern of the LiFeAs single crystal aligned along the (001) plane. The Miller indices of each peak are represented in the figure. The inset shows a photograph of a piece of grown crystal. One grid in the photograph represents 1 mm.

Fig. 2 (Color online) (a) Temperature dependent resistivity of the LiFeAs crystal along the $ab$-plane around superconducting transition. The inset presents the resistivity in a wide temperature region from 3 to 300 K. (b) Temperature dependence of DC magnetic susceptibility. Zero-field-cooled (ZFC) and field-cooled (FC) measurements were performed with $H$ = 10 Oe applied along the $ab$-plane. $T_c$ is realized as a temperature where the demagnetization drop starts to appear, which is shown as a dashed line in the inset. (c) $\Delta C_p/T$ near $T_c$ at zero magnetic field. The thermodynamic $T_c \approx 16.8$ K and a jump $\Delta C_p/T \approx 20.0$ mJ/mol·K$^2$ are estimated through the linear extrapolations (dashed line). The temperature dependence of $C_p/T$ in a wide temperature range is presented in the inset. The solid line in the inset is the estimated normal state contribution as explained in the text.

Fig. 3 (Color online) Resistivity of LiFeAs single crystal measured at different magnetic fields. The magnetic field was applied (a) along the $c$-axis and (b) the $ab$-plane.

Fig. 4 (Color online) Phase diagram of $H_{c2}$ vs. temperature. $T_c$ for each magnetic field was determined by the criterion of 50 % of the normal state resistivity at $T_c$. The inset presents temperature dependence of the superconducting anisotropy $\gamma$, obtained upon interpolation of the $H_{c2}(T)$ curves.

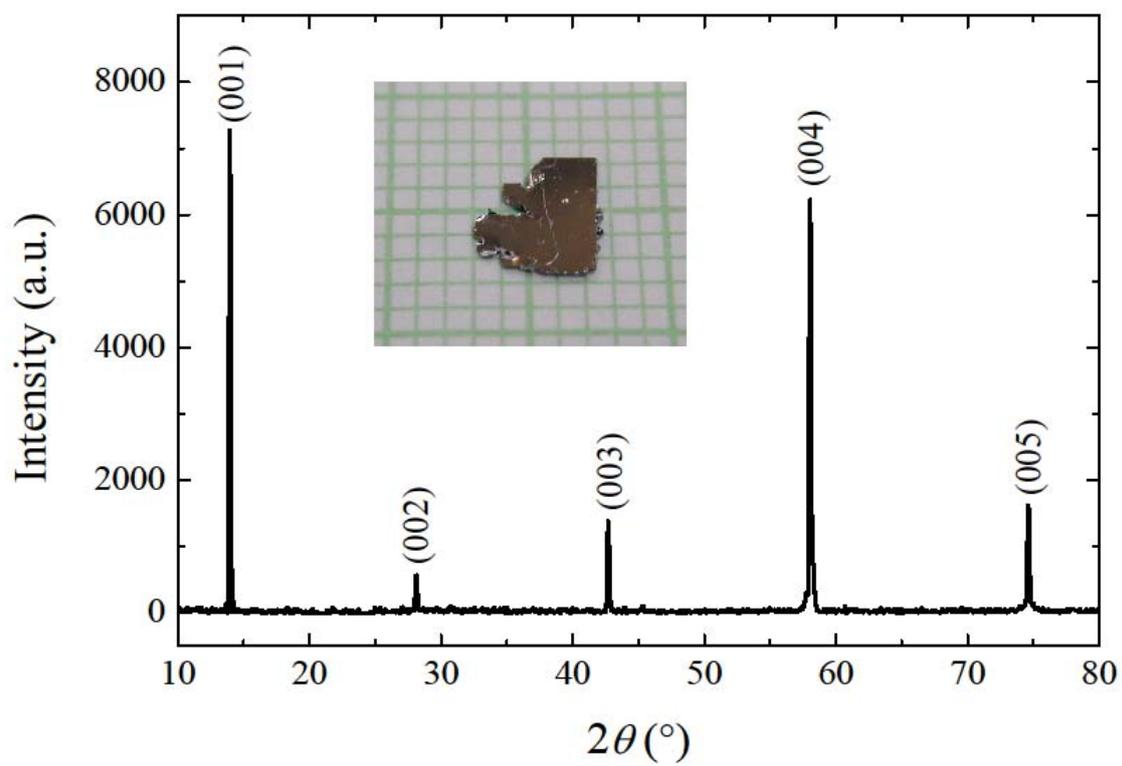

**Fig. 1 Bumsung Lee *et al*.**

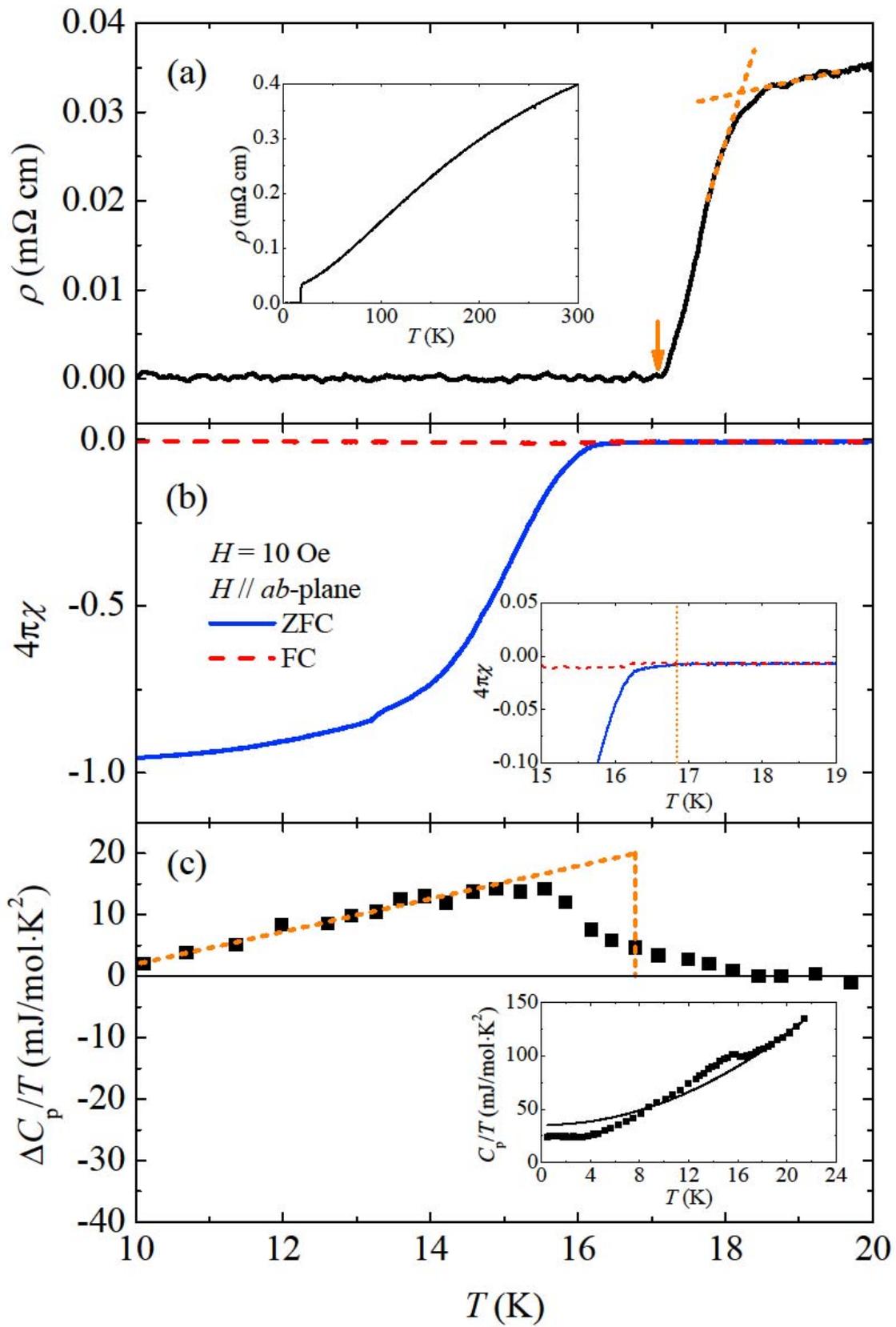

**Fig. 2 Bumsung Lee *et al*.**

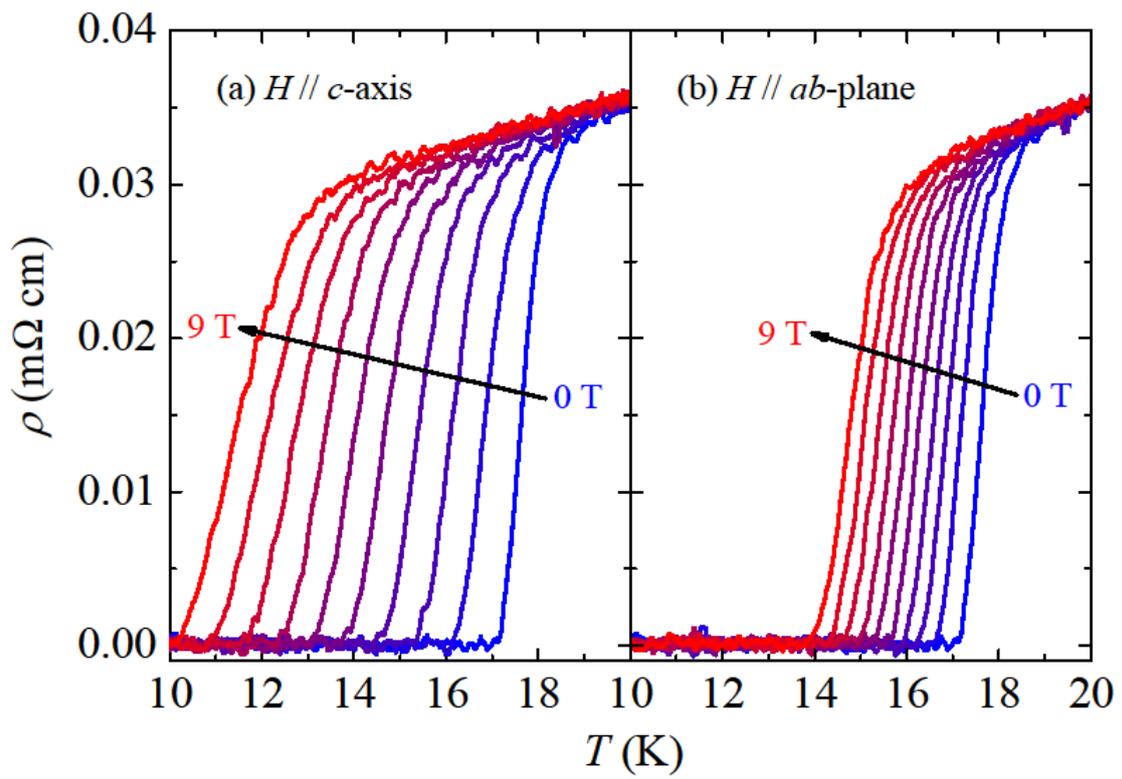

Fig. 3 Bumsung Lee *et al*.

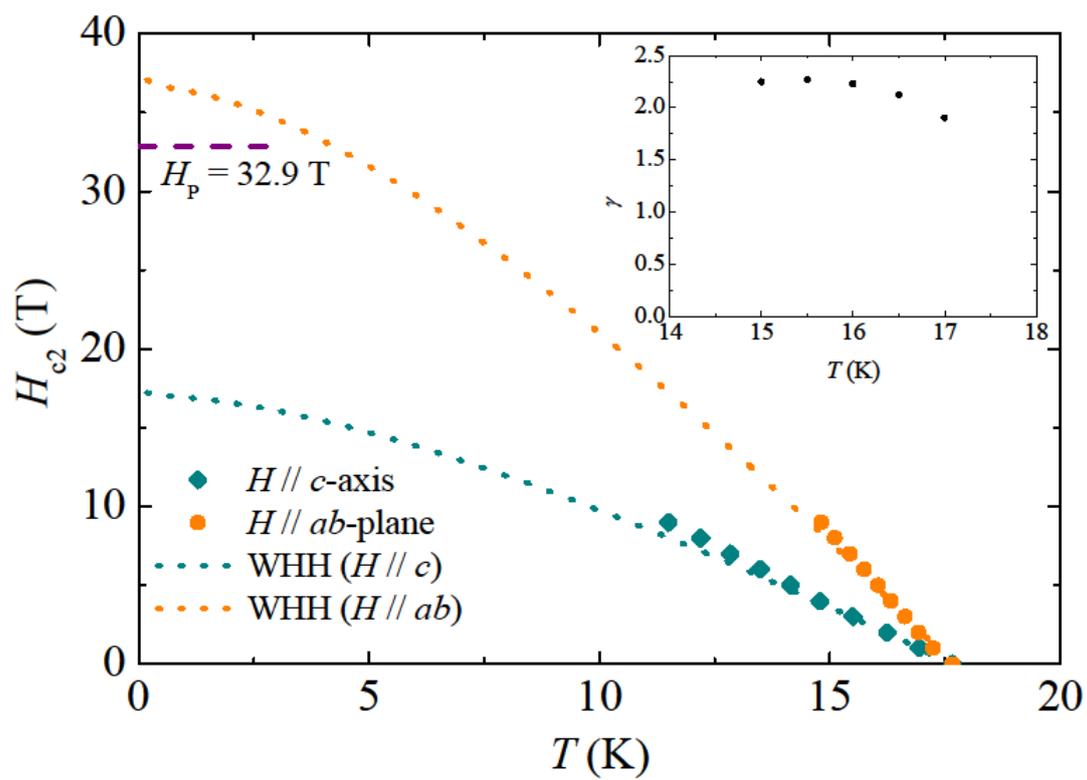

Fig. 4 Bumsung Lee *et al*.